


\input lanlmac

\input epsf

\newcount\figno
\figno=0
\def\fig#1#2#3{
\par\begingroup\parindent=0pt\leftskip=1cm\rightskip=1cm\parindent=0pt
\baselineskip=11pt
\global\advance\figno by 1
\midinsert
\epsfxsize=#3
\centerline{\epsfbox{#2}}
\vskip 12pt
\centerline{{\bf Fig. \the\figno:~~} #1}\par
\endinsert\endgroup\par
}
\def\figlabel#1{\xdef#1{\the\figno}}


\def\th{\theta}

\def\vep{\varepsilon}

\def\Z{{\bf Z}}

\def\Tr{{\rm Tr}}
\def\hf{{1\over 2}}
\def\qu{{1\over 4}}

\def\o{\over}
\def\til#1{\widetilde{#1}}
\def\si{\sigma}
\def\Si{\Sigma}
\def\b#1{\overline{#1}}
\def\del{\partial}

\def\bra{\langle}
\def\ket{\rangle}
\def\lf{\left}
\def\ri{\right}
\def\riya{\rightarrow}

\def\lrya{\leftrightarrow}

\def\la{\lambda}
\def\La{\Lambda}
\def\h#1{\widehat{#1}}

\def\ga{\gamma}

\def\al{\alpha}
\def\om{\omega}

\def\tens{\otimes}

\def\dag{\dagger}
\def\rt#1{\sqrt{#1}}

\def\sitarel#1#2{\mathrel{\mathop{\kern0pt #1}\limits_{#2}}}

\def\cob{\delta}

\def\nor{{}^\circ \hskip-0.16cm {}_{\circ}}
\def\bul{{}^\bullet \hskip-0.16cm {}_{\bullet}}


\lref\Sbrane{
M.~Gutperle and A.~Strominger,
``Spacelike branes,''
JHEP {\bf 0204}, 018 (2002)
[hep-th/0202210].
}

\lref\SenNU{
A.~Sen,
``Rolling tachyon,''
JHEP {\bf 0204}, 048 (2002)
[hep-th/0203211].
}

\lref\SenTM{
A.~Sen,
``Tachyon matter,''
JHEP {\bf 0207}, 065 (2002)
[hep-th/0203265].
}

\lref\SenT{
A.~Sen,
``Field theory of tachyon matter,''
Mod.\ Phys.\ Lett.\ A {\bf 17}, 1797 (2002)
[hep-th/0204143].
}

\lref\SenTime{
A.~Sen,
``Time and tachyon,''
[hep-th/0209122].
}

\lref\SenBI{
A.~Sen,
``Dirac-Born-Infeld action on the tachyon kink and vortex,''
[hep-th/0303057].
}

\lref\Sthermo{
A.~Maloney, A.~Strominger and X.~Yin,
``S-brane thermodynamics,''
[hep-th/0302146].
}

\lref\Lambert{
N.~D.~Lambert and I.~Sachs,
``Tachyon dynamics and the effective action approximation,''
Phys.\ Rev.\ D {\bf 67}, 026005 (2003)
[hep-th/0208217].
}

\lref\Spha{
J.~A.~Harvey, P.~Horava and P.~Kraus,
``D-sphalerons and the topology of string configuration space,''
JHEP {\bf 0003}, 021 (2000)
[hep-th/0001143].
}
\lref\LLM{
N.~Lambert, H.~Liu and J.~Maldacena,
``Closed strings from decaying D-branes,''
[hep-th/0303139].
}

\lref\SenVV{
A.~Sen,
``Time evolution in open string theory,''
JHEP {\bf 0210}, 003 (2002)
[hep-th/0207105].
}

\lref\MukhopadhyayEN{
P.~Mukhopadhyay and A.~Sen,
``Decay of unstable D-branes with electric field,''
JHEP {\bf 0211}, 047 (2002)
[hep-th/0208142].
}
\lref\StromingerPC{
A.~Strominger,
``Open string creation by S-branes,''
[hep-th/0209090].
}

\lref\ChenFP{
B.~Chen, M.~Li and F.~L.~Lin,
``Gravitational radiation of rolling tachyon,''
JHEP {\bf 0211}, 050 (2002)
[hep-th/0209222].
}

\lref\ReyXS{
S.~J.~Rey and S.~Sugimoto,
``Rolling tachyon with electric and magnetic fields: T-duality approach,''
Phys.\ Rev.\ D {\bf 67}, 086008 (2003)
[hep-th/0301049].
}

\lref\Larsen{
F.~Larsen, A.~Naqvi and S.~Terashima,
``Rolling tachyons and decaying branes,''
JHEP {\bf 0302}, 039 (2003)
[hep-th/0212248].
}

\lref\OkudaYD{
T.~Okuda and S.~Sugimoto,
``Coupling of rolling tachyon to closed strings,''
Nucl.\ Phys.\ B {\bf 647}, 101 (2002)
[hep-th/0208196].
}

\lref\ReyZJ{
S.~J.~Rey and S.~Sugimoto,
``Rolling of modulated tachyon with gauge flux and emergent fundamental  string,'' Phys.\ Rev.\ D {\bf 68}, 026003 (2003)
[hep-th/0303133]
}

\lref\Gutperle{
M.~Gutperle and A.~Strominger,
``Timelike Boundary Liouville Theory,''
Phys.\ Rev.\ D {\bf 67}, 126002 (2003)
[hep-th/0301038].
}

\lref\GaiottoRM{
D.~Gaiotto, N.~Itzhaki and L.~Rastelli,
``Closed strings as imaginary D-branes,''
[hep-th/0304192].
}

\lref\NeilFinn{
N.~R.~Constable and F.~Larsen,
``The rolling tachyon as a matrix model,''
JHEP {\bf 0306}, 017 (2003)
[hep-th/0305177].
}

\lref\Sugawara{
Y.~Sugawara,
``Thermal Partition Functions for S-branes,''
JHEP {\bf 0308}, 008 (2003)
[hep-th/0307034].
}

\lref\SenIV{
A.~Sen,
``Open-closed duality: Lessons from matrix model,''
[hep-th/0308068].
}

\lref\FateevIK{
V.~Fateev, A.~B.~Zamolodchikov and A.~B.~Zamolodchikov,
``Boundary Liouville field theory. I: 
Boundary state and boundary  two-point function,''
[hep-th/0001012].
}

\lref\GaberdielXM{
M.~R.~Gaberdiel, A.~Recknagel and G.~M.~Watts,
``The conformal boundary states for SU(2) at level 1,''
Nucl.\ Phys.\ B {\bf 626}, 344 (2002)
[hep-th/0108102].
}

\lref\StromingerFN{
A.~Strominger and T.~Takayanagi,
``Correlators in timelike bulk Liouville theory,''
[hep-th/0303221].
}

\lref\SchomerusVV{
V.~Schomerus,
``Rolling tachyons from Liouville theory,''
[hep-th/0306026].
}

\lref\SenXS{
A.~Sen,
``Open-closed duality at tree level,''
[hep-th/0306137].
}

\lref\CallanUB{
C.~G.~Callan, I.~R.~Klebanov, A.~W.~Ludwig and J.~M.~Maldacena,
``Exact solution of a boundary conformal field theory,''
Nucl.\ Phys.\ B {\bf 422}, 417 (1994)
[hep-th/9402113].
}

\lref\KarczmarekXM{
J.~L.~Karczmarek, H.~Liu, J.~Maldacena and A.~Strominger,
``UV finite brane decay,''
[hep-th/0306132].
}

\lref\McGreevyEP{
J.~McGreevy, J.~Teschner and H.~Verlinde,
``Classical and quantum D-branes in 2D string theory,''
[hep-th/0305194].
}

\lref\DouglasWY{
M.~R.~Douglas,
``Conformal field theory techniques in large N Yang-Mills theory,''
[hep-th/9311130].
}

\lref\CordesFC{
S.~Cordes, G.~W.~Moore and S.~Ramgoolam,
``Lectures on 2-d Yang-Mills theory, 
equivariant cohomology and topological field theories,''
Nucl.\ Phys.\ Proc.\ Suppl.\  {\bf 41}, 184 (1995)
[hep-th/9411210].
}

\lref\GrossHE{
D.~J.~Gross and E.~Witten,
``Possible Third Order Phase Transition In The Large N Lattice Gauge Theory,''
Phys.\ Rev.\ D {\bf 21}, 446 (1980).
}

\lref\KazakovPM{
V.~Kazakov, I.~K.~Kostov and D.~Kutasov,
``A matrix model for the two-dimensional black hole,''
Nucl.\ Phys.\ B {\bf 622}, 141 (2002)
[hep-th/0101011].
}

\lref\VafaQF{
C.~Vafa,
``Brane/anti-brane systems and U(N$|$M) supergroup,''
[hep-th/0101218].
}

\lref\GoulianQR{
M.~Goulian and M.~Li,
``Correlation Functions In Liouville Theory,''
Phys.\ Rev.\ Lett.\  {\bf 66}, 2051 (1991).
}

\lref\DouglasEX{
M.~R.~Douglas,
``Chern-Simons-Witten theory as a topological Fermi liquid,''
[hep-th/9403119].
}

\lref\Jinzenji{
M.~Jinzenji and T.~Sasaki,
``N = 4 supersymmetric Yang-Mills theory on orbifold $T^4/Z_2$:  
Higher rank case,''
JHEP {\bf 0112}, 002 (2001)
[hep-th/0109159].
}

\lref\Macdonald{
I. G. Macdonald, ``Affine root systems and Dedekind's $\eta$-function,''
Invent. Math. {\bf 15}, 91 (1972).  
}

\lref\Kac{
V. G. Kac, ``Infinite-dimensional Lie algebras,''
3rd edition, Cambridge University Press, 1990.
}

\lref\Polchinski{
J. Polchinski, ``String Theory,'' volume I, Cambridge University Press, 1998. 
}

\lref\Macbook{
I. G. Macdonald, ``Symmetric Functions and Hall Polynomials,''
2nd edition, Oxford University Press, 1995.
}

\lref\Wakimoto{
V.~G.~Kac and M.~Wakimoto,
``Integrable Highest Weight Modules Over Affine 
Superalgebras And Number Theory,''
[hep-th/9407057].
}

\lref\Zagier{
D. Zagier,
`` A proof of the Kac-Wakimoto affine denominator formula 
for the strange series,'' Math. Res. Lett. {\bf 7}, 597 (2000). 
}

\lref\MatsuoCJ{
Y.~Matsuo,
``Character Formula Of $c < 1$ 
Unitary Representation Of N=2 Superconformal Algebra,''
Prog.\ Theor.\ Phys.\  {\bf 77}, 793 (1987).
}

\Title{
\vbox{\hbox{EFI-03-41}
      \hbox{hep-th/0308172}}
}
{Comments on Half S-Branes}

\vskip .2in

\centerline{Kazumi Okuyama}

\vskip .2in

\centerline{ Enrico Fermi Institute, University of Chicago} 
\centerline{ 5640 S. Ellis Ave., Chicago IL 60637, USA}
\centerline{\tt kazumi@theory.uchicago.edu}

\vskip 3cm
\noindent

Following hep-th/0305177,
we write the boundary state of half S-brane in bosonic string theory
as a grand canonical partition function of a unitary matrix model.
From this representation, 
it follows that the annulus amplitude can be written as a 
grand canonical partition function
of a unitary two-matrix model.
We also show that the contribution of the exponentially growing couplings to the
timelike oscillators can be resummed in a certain annulus amplitude. 
 
\Date{August 2003}

\vfill
\vfill

\newsec{Introduction}
The decay of unstable D-brane is an interesting laboratory to study 
various properties of string theory in time-dependent backgrounds 
\refs{\Sbrane\SenNU\SenTM\SenT\SenVV\SenTime\MukhopadhyayEN\StromingerPC\ChenFP\Larsen\Gutperle\Sthermo\OkudaYD\ReyXS\ReyZJ\LLM\GaiottoRM\NeilFinn\KarczmarekXM{--}\Sugawara}.
Moreover, the study of unstable D-brane will 
give us a new perspective to the 
open/closed duality \refs{\SenXS,\SenIV}. 
In this paper, we consider the decay of D-brane in bosonic string theory
described by the timelike boundary Liouville theory:  
\eqn\STBL{
S_{\rm TBL}=-{1\o2\pi}\int_{\Si}\del X^0\b{\del}X^0+\la\oint_{\del\Si}e^{X^0},
}
which is also referred to as the half S-brane.
Throughout this paper, we set $\al'=1$.
This negative norm boson theory can be defined by an analytic continuation 
of the spacelike boundary Liouville theory. 

There are two different approaches to define this analytic continuation.
In the first approach, we
start with the spacelike Liouville theory with background
charge $Q=b+b^{-1}$ and take the limit $b\riya i$ \Gutperle.
In this approach, the half S-brane boundary state is 
obtained by the analytic continuation of
the boundary state in the spacelike Liouville theory,
which describes the D-brane extended in the Liouville direction \FateevIK,
\eqn\BFZZ{
|B_{\rm LFZZ}\ket=\int_{-\infty}^\infty dE~
\la^{-iE}{\pi\o\sinh \pi E}|E\,\ket\ket.
}
We call this state the ``Lorentzian FZZ state.'' This is expanded
in the Ishibashi state $|E\,\ket\ket$
built on the bulk primary operator $e^{iEX^0}$.
In the second approach, which is adopted 
in the Sen's original paper \SenNU,
we perform an inverse Wick rotation of the boundary state of
a spacelike $c=1$ boson $X$ with the boundary interaction $e^{iX}$ 
\refs{\CallanUB,\GaberdielXM,\Larsen,\Gutperle}: 
\eqn\Bdisc{
|B\ket=\sum_{j}\sum_{m\geq0}^j\lf(\matrix{j+m\cr 2m}\ri)(i\la)^{2m}
|j,m,m\ket\ket.
}
In this case, the boundary state is expanded in terms of the
Ishibashi state built on the discrete states.
Note that in these two approaches we analytically continue from the 
spacelike CFTs with different central charges: $c\not=1$ for \BFZZ\ and
$c=1$ for \Bdisc.

It is important to notice that
the resulting Lorentzian boundary states \BFZZ\ and \Bdisc\ are {\it different}.
In particular, the exponentially growing couplings to the
timelike oscillators \OkudaYD\ in \Bdisc\ are not included in \BFZZ.
Mathematically, this reflects the fact that the analytic continuation
to define \STBL\ is not unique.
Physically, as discussed in \refs{\Gutperle,\LLM,\KarczmarekXM,\StromingerFN,
\SchomerusVV,\SenXS}, in time-dependent backgrounds the worldsheet CFT 
has ambiguities
corresponding to the choices of incoming state, or vacuum.
Therefore, we should regard these two different 
boundary states \BFZZ\ and \Bdisc\ as a result of the two different 
choices of the vacuum.

Most of the computations of Lorentzian annulus amplitude
considered in the literature are based on the
Lorentzian FZZ state \BFZZ.
Therefore, it is natural to ask what happens if we use \Bdisc\ in the
computation of annulus amplitude.
However, the exponentially growing couplings to the massive modes found in 
\OkudaYD\ make this very difficult, 
both technically and conceptually.
It was shown \OkudaYD\ that the coupling to the timelike oscillator at
a {\it fixed} level $n\geq2$ in \Bdisc\ grows exponentially at late times,
and we cannot resum these terms as opposed to the situation at level 0 and 1.
However, there is a possibility to perform a resummation
over the terms at different levels.
The recent result of \NeilFinn\ will be potentially useful
for this purpose, because it provides us an efficient way to
compute the coefficient of \Bdisc\ in terms of the unitary matrix
integral. In this paper, we will show that 
the contribution of those exponentially growing terms
can be resummed in a certain annulus amplitude.

This paper is organized as follows.
In section 2, after reviewing the result of \NeilFinn, we 
write the boundary state \Bdisc\ as a grand canonical partition function 
of a unitary matrix model.
We also obtain a useful decomposition of $|B\ket$, in which
the first term corresponds to $|B_{\rm LFZZ}\ket$ and the second
term represents the exponentially growing coupling to 
the timelike oscillators \OkudaYD.
In section 3, we write the annulus amplitude between two $|B\ket$'s
as a grand canonical partition function of a unitary two-matrix model,
and present some examples of the matrix integral.
In section 4, we consider the annulus amplitude between 
$|B\ket$ and the first term in the decomposition of $|B\ket$ obtained in
section 2. We show that in this amplitude we can perform a resummation of the
exponentially growing timelike oscillators.
Section 5 is devoted to discussions. In appendix A, we give a proof of some
identity used in section 4.

\newsec{Half S-brane Boundary State}
In this section, we consider the boundary state $|B\ket$
of timelike boundary Liouville theory \STBL.
Following \NeilFinn, we write $|B\ket$ as a grand canonical partition function 
of a unitary matrix model.

In this paper, we use the following 
notation for a partition $\mu=(n_1^{N_1}\cdots n_k^{N_k})$:
\eqn\partitdef{
|\mu|=\sum_i n_iN_i,\quad \ell(\mu)=\sum_iN_i,\quad
z_\mu=\prod_i n_i^{N_i}N_i!,
}
{\it i.e.}, $|\mu|$, $\ell(\mu)$, and $z_{\mu}$ denote the
total number, the length, and the norm of the partition $\mu$, respectively.
$n_i$'s are strictly increasing: $n_1<\cdots<n_k$.

\subsec{Boundary State and Unitary Matrix Integral}
To write the boundary state as a matrix integral,
we first review the result of \NeilFinn.
It was shown in \NeilFinn\ that
the disk one-point function 
$\bra\nor V^{(\si;\tilde{\si})}\nor\ket$ of the operator
\eqn\Vsiop{
V^{(\si;\tilde{\si})}={1\o \rt{z_{\si}z_{\tilde{\si}}}}
\prod_{i,j}\lf({\rt{2}\o(n_i-1)!}\del^{n_i}X^0\ri)^{N_i}
\lf({\rt{2}\o(\tilde{n}_j-1)!}\b{\del}^{\tilde{n}_j}X^0\ri)^{\til{N}_j},
}
labeled by $\si=(n_i^{N_i})$ and $\til{\si}=(\tilde{n}_i^{\tilde{N}_i})$,
is given by the unitary matrix integral
\eqn\AtoI{
\bra\nor V^{(\si;\tilde{\si})}\nor\ket
=\sum_{N=0}^{\infty}(-\til{\la})^N{\cal N}'^{(\si;\tilde{\si})}
\int_{U(N)}dU\,
\Tr\,U_{\si}\Tr\,U^{-1}_{\tilde{\si}}~;\quad
{\cal N}'^{(\si;\tilde{\si})}
=\rt{{2^{\ell(\si)}\o z_{\si}}{2^{\ell(\tilde{\si})}\o z_{\tilde{\si}}}}.
}
The measure $dU$ is normalized so that $\int_{U(N)}dU=1$.
This remarkable relation \AtoI\ was obtained  by expanding the boundary 
interaction in terms of
the ``time-dependent coupling''\foot{In \NeilFinn, $\la$ and $\til{\la}$ were
denoted as $g$ and $\til{g}$, respectively.}
\eqn\tildelamblda{
\til{\la}=\la\,e^{x^0},
}
and then performing the path integral over the 
non-zero modes of $X^0$. $x^0$ in $\til{\la}$ is the zero-mode
of the worldsheet boson $X^0$.
The rank $N$ of group $U(N)$ is the number of insertions of tachyon operator,
and the $N$ eigenvalues of $U$ describes the position of
the $N$ tachyon operators on the unit circle, which is the boundary of the disk
worldsheet.
The operator $V^{(\si;\tilde{\si})}$ in \Vsiop\ corresponds to the state
\eqn\sigtilsig{
|\sigma,\tilde{\sigma}\rangle={1\o\rt{z_\si z_{\tilde{\si}}}}
\alpha_{-\sigma}
\til{\alpha}_{-\tilde{\sigma}}|0\rangle,
} 
under the mapping rule
\eqn\opstate{
\al_{-n}\lrya {\rt{2}i\o(n-1)!}\del^{n}X= {\rt{2}\o(n-1)!}\del^{n}X^0.
}
In \AtoI\ and \sigtilsig, we used the abbreviation
\eqn\alsig{
\al_{-\si}=\prod_i\al_{-n_i}^{N_i},\quad
\Tr\,U_\si=\prod_i\lf(\Tr\,U^{n_i}\ri)^{N_i}.
}
The normalization of $|\si,\tilde{\si}\ket$ is chosen so that
$\bra \si,\tilde{\si}|\si',\tilde{\si}'\ket
=\cob_{\si,\si'}\cob_{\tilde{\si},\tilde{\si}'}$.
In \sigtilsig, 
the oscillators $\al_n$ should be inverse Wick rotated $\al_n\riya-i\al_n^0$
when we talk about the boundary state of the timelike Liouville theory.
But we continue using the Euclidean notation $\al_n$ in order to
avoid the appearance of $i$'s in equations.

The ordering $\nor~\nor$ appearing in $\bra\nor V^{(\si;\tilde{\si})}\nor\ket$ 
is the boundary
normal ordering defined by the Green function
on the disk including the effect of the image charge.
On the other hand, the overlap between the boundary state $|B\ket$
and the state $|\si,\tilde{\si}\ket$ 
is given by the disk amplitude 
$\bra\bul V^{(\si;\tilde{\si})}\bul\ket$
computed with respect to the bulk normal ordering $\bul~\bul$.
These two normal orderings are related by 
\eqn\normorder{
\bul X^0(z,\b{z})X^0(w,\b{w})\bul
=\nor X^0(z,\b{z})X^0(w,\b{w})\nor+\log|z\b{w}-1|.
}
The relation between the two amplitude $\bra\bul V^{(\si;\tilde{\si})}\bul\ket$
and $\bra\nor V^{(\si;\tilde{\si})}\nor\ket$ can be written in a compact form 
in terms of the states 
\eqn\BandA{
|B\ket
=\sum_{\sigma,\tilde{\sigma}}\bra\bul V^{(\si;\tilde{\si})}\bul\ket
|\sigma,\tilde{\sigma}\rangle,\quad
| A\ket
=\sum_{\sigma,\tilde{\sigma}}\bra\nor V^{(\si;\tilde{\si})}\nor\ket
|\sigma,\tilde{\sigma}\rangle.
}
By noticing the relation
\eqn\normoedosci{
\bul{\rt{2}\o(n-1)!}\del^{n}X^0(0){\rt{2}\o(m-1)!}\del^{m}X^0(0)\bul
=\nor{\rt{2}\o(n-1)!}\del^{n}X^0(0){\rt{2}\o(m-1)!}\del^{m}X^0(0)
\nor-n\cob_{n,m},
}
and the correspondence rule \opstate,
it is easy to see that $|B\rangle$ and $|A\rangle$ are related by\foot{
See {\it e.g.} eq.(2.7.14) in \Polchinski.
This derivation is due to F. Larsen. We would like to thank 
N. Constable and F. Larsen for discussion on this point.}
\eqn\BexpA{
|B\rangle=\exp\left(-\sum_{n=1}^\infty{1\over n}\alpha_{-n}
\til{\alpha}_{-n}\right)|A\rangle.
}

Plugging \AtoI\ into \BandA, 
and performing the summation over $\sigma$ and $\tilde{\sigma}$,
$|A\ket$ is written as
\eqn\AasgrandP{
|A\rangle=\sum_{N=0}^\infty(-\til{\la})^N\int_{U(N)}dU
\exp\left[\sum_{n=1}^\infty{\sqrt{2}\over n}\Big({\rm Tr}\,U^n\alpha_{-n}
+{\rm Tr}\,U^{- n}\til{\alpha}_{-n}\Big)\right]|0\rangle.
}
Finally,
using the relation \BexpA\ between $|A\ket$ and $|B\ket$, 
we arrive at the main result of this section:
\eqn\BasgrandP{\eqalign{
|B\rangle&=\sum_{N=0}^\infty(-\til{\la})^N\int_{U(N)}dU
\exp\left[\sum_{n=1}^\infty{1\over n}\Big(-\alpha_{-n}\til{\alpha}_{-n}+
\sqrt{2}{\rm Tr}\,U^n\alpha_{-n}
+\sqrt{2}{\rm Tr}\,U^{- n}\til{\alpha}_{-n}\Big)\right]|0\rangle
\cr
&\equiv\sum_{N=0}^\infty(-\til{\la})^N\int_{U(N)}dU|B_U\rangle,
}}
where $|B_U\rangle$ is the coherent state obeying
\eqn\BUbc{
(\alpha_{n}+\til{\alpha}_{-n})|B_U\rangle=\sqrt{2}{\rm Tr}\,U^n|B_U\rangle.
}
Note that in the Euclidean theory 
the momentum is proportional to the rank of $U(N)$
\eqn\zeroalasN{
\al_0=\til{\al}_0=\rt{{\al'\o2}}p={\rt{2}\o2}\Tr\,{\bf 1}={1\o\rt{2}}N.
}
To summarize, the boundary state of timelike Liouville theory
is the grand canonical partition function 
of the unitary matrix integral \BasgrandP.
Here, the zero-mode $x^0$, or time, 
plays the role of the chemical potential for $N$.
\foot{We would like to thank N. Constable for pointing out this
interpretation.}

Another useful expression of $|B\rangle$ can be obtained 
by expanding the matrix integral in terms of the characters $\chi_{Y}(\si)$ 
of symmetric group. 
Using the Frobenius relation, the matrix integral 
is written as 
\eqn\INchi{
\int_{U(N)}dU\Tr\,U_{\si}\Tr\,U^{-1}_{\tilde{\si}}=
\sum_{\ell(Y)\leq N}\chi_{Y}(\sigma)\chi_{Y}(\tilde{\sigma})
=z_{\si}\delta_{\sigma,\tilde{\sigma}}
-\sum_{\ell(Y)> N}\chi_{Y}(\sigma)\chi_{Y}(\tilde{\sigma}).
}
Here we used the completeness relation $\sum_{Y}\chi_{Y}(\si)
\chi_{Y}(\tilde{\si})=z_{\si}\cob_{\si,\tilde{\si}}$.
Plugging \INchi\ into \AtoI,
$|A\rangle$ is rewritten as
\eqn\Aexpchi{\eqalign{
|A\rangle
&=\sum_{N=0}^\infty(-\til{\la})^N\sum_{\si,\tilde{\si}}
{\cal N}'^{(\si,\tilde{\si})}\sum_{\ell(Y)\leq N}\chi_Y(\si)
\chi_{Y}(\tilde{\si})|\si,\tilde{\si}\ket\cr
&= f\exp\left(2\sum_{n=1}^\infty{1\over n}
\alpha_{-n}\tilde{\alpha}_{-n}\right)|0\rangle
-\sum_{N=0}^\infty(-\til{\la})^N
\sum_{\sigma,\tilde{\sigma}}{\cal N}'^{(\sigma,\tilde{\sigma})}
\sum_{\ell(Y)> N}\chi_{Y}(\sigma)\chi_{Y}(\tilde{\sigma})
|\sigma,\tilde{\sigma}\rangle \cr
&= (f-1)\exp\left(2\sum_{n=1}^\infty{1\over n}
\alpha_{-n}\tilde{\alpha}_{-n}\right)|0\rangle
+|0\rangle \cr
& \hskip10mm-\sum_{N=1}^\infty(-\til{\la})^N
\sum_{\sigma,\tilde{\sigma}}{\cal N}'^{(\sigma,\tilde{\sigma})}
\sum_{\ell(Y)> N}\chi_{Y}(\sigma)\chi_{Y}(\tilde{\sigma})
|\sigma,\tilde{\sigma}\rangle.
}}
In the third line, we rewrote the $N=0$ term by using
\eqn\chizero{
\sum_{\ell(Y)> 0}\chi_{Y}(\sigma)\chi_{Y}(\tilde{\sigma})
=\delta_{\sigma,\tilde{\sigma}}z_{\si}
-\delta_{\sigma,0}\delta_{\tilde{\sigma},0}.
}
$f$ in \Aexpchi\ is given by
\eqn\fxzero{
f={1\o1+\til{\la}}={1\o 1+\la\,e^{x^0}}.
}
From \BexpA\ and \Aexpchi, we obtain the following decomposition of
the boundary state:
\eqn\Bdecomp{
|B\rangle=f|D\ket+|B'_0\ket=
(f-1)|D\rangle+|N\rangle+|B'_1\rangle.
}
Here, $|D\rangle$ and $|N\rangle$ are the usual Dirichlet and Neumann
states
\eqn\DandNdef{
|D\rangle=\exp\left(\sum_{n=1}^\infty{1\over n}
\alpha_{-n}\tilde{\alpha}_{-n}\right)|0\rangle,\quad
|N\rangle=\exp\left(-\sum_{n=1}^\infty{1\over n}
\alpha_{-n}\tilde{\alpha}_{-n}\right)|0\rangle,
}
and $|B'_{k}\rangle~(k=0,1)$ is the extra piece 
\eqn\Bextra{
|B'_{k}\rangle=-\sum_{N=k}^\infty(-\til{\la})^N
\sum_{\sigma,\tilde{\sigma}}{\cal N}'^{(\sigma,\tilde{\sigma})}
\sum_{\ell(Y)> N}\chi_{Y}(\sigma)\chi_{Y}(\tilde{\sigma})
\exp\left(-\sum_{n=1}^\infty{1\over n}
\alpha_{-n}\tilde{\alpha}_{-n}\right)|\sigma,\tilde{\sigma}\rangle.
}
The exponentially growing terms found by Okuda and Sugimoto \OkudaYD\
are contained solely in $|B_1'\ket$.
Note also that $|B\rangle$ reduces to the Neumann state 
$|N\rangle$ when $\la=0$, as expected. 

From \INchi, one can see that
the growing terms in $|B_0'\ket$ are coming from the failure of 
the completeness of $\chi_{Y}(\si)$ due to the constraint $\ell(Y)\leq N$.
We expect that
the constraint $\ell(N)\leq N$ becomes
irrelevant in the large $N$ limit.
In fact, it is well known that in the large $N$ limit the
traces of unitary matrix behave as the free Gaussian variables \DouglasWY\foot{
Since we compute the matrix integral by expanding the exponential 
\BasgrandP,
our large $N$ limit corresponds to the strong coupling phase of
the Gross-Witten phase transition \GrossHE,
in which the eigenvalue distribution is uniform on the unit circle. 
We will not discuss the interpretation
of the weak coupling phase in the rolling tachyon boundary state.
} 
\eqn\largeN{
\lim_{N\riya\infty}\int_{U(N)}dU\,F(\Tr U^n,\Tr U^{-n})=
\int\prod_{m=1}^\infty d\mu_md\b{\mu}_me^{-\mu_m\b{\mu}_m}
\,F(\rt{n}\mu_n,\rt{n}\b{\mu}_n).
}
Using this relation, it is easy to see that 
\eqn\largeNBN{
\lim_{N\riya\infty}\int_{U(N)}dU|B_U\ket=|D\ket.
}
If we replace all the finite $N$ integral in \BasgrandP\ by the large $N$ 
value \largeNBN, we get 
\eqn\BsimlargeN{
|B\ket\sim \sum_{N=0}^\infty(-\til{\la})^N|D\ket=f|D\ket.
}
Therefore, the first term in the decomposition \Bdecomp\
can be thought of as the large $N$ result and 
the extra piece $|B_0'\ket$ corresponds to
the finite $N$ effect.

Before concluding this subsection, 
let us comment on our treatment of the zero-mode $x^0$.
In the boundary state $|B\ket$, 
the boundary interaction $\int_{\del\Si} e^{X^0}$
acts on the Neumann state at $\tau=0$ in the closed string channel.
Since the conjugate momentum of $x^0$ does not appear in $X^0(\tau=0)$ 
\eqn\Xpzero{
X^0(\tau=0)=x^0+p^0\tau+{\rm oscillators}\Big|_{\tau=0}=
x^0+{\rm oscillators},
}
the zero-mode $x^0$ can be treated as if it is a c-number.
For instance, $f$ in \fxzero\ should be understood as the 
eigenvalue in the coordinate basis \OkudaYD
\eqn\fasstate{
f\lrya \int dx^0 f(x^0)|x^0\ket.
}
It is interesting to compare this treatment of zero-mode with that
in the spacelike Liouville theory. In the spacelike theory, it is sometimes
useful to integrate out the zero-mode first and then integrate 
the non-zero modes \GoulianQR. On the other hand, in the timelike theory
it turned out to be useful to integrate the zero-mode last and treat it as a
part of the coupling \refs{\Larsen,\NeilFinn}.

\subsec{Relation to the $c=1$ Discrete States}
It was argued in \NeilFinn\ that the boundary state
\Bdisc\ written in terms of the discrete states and 
the matrix integral representation of the boundary state \BasgrandP\
are equivalent. This was checked up to the level 3 of oscillator number 
\NeilFinn. 
It is not so easy to see which terms in \BasgrandP\ correspond to 
the Virasoro Ishibashi state $|j,m,m\ket\ket$.
However, in some cases we can identify the primary states $|j,m,m\ket$
in the matrix representation of $|B\ket $\BasgrandP.
For instance, the level 0 term $f|0\ket$
is obtained by summing over the primary states with $m=j$ 
\refs{\SenNU,\Larsen,\Gutperle}. 
We can also identify the primary state with $m=j-1$.
The explicit form of $|B_1'\ket$ up to level 3 is
\eqn\Bonepri{\eqalign{
|B_1'\ket=&~\til{\la}\,\Bigg[\hf(\al_{-2}-\rt{2}\al_{-1}^2)
(\til{\al}_{-2}-\rt{2}\til{\al}_{-1}^2) \cr
&~\,+\Big({2\o3}\al_{-3}-{1\o\rt{2}}\al_{-2}\al_{-1}-{1\o3}\al_{-1}^3\Big)
\Big({2\o3}\til{\al}_{-3}-{1\o\rt{2}}\til{\al}_{-2}\til{\al}_{-1}
-{1\o3}\til{\al}_{-1}^3\Big)+\cdots\Bigg]|0\ket\cr
-&\til{\la}^2\lf[
\Big({\rt{2}\o3}\al_{-3}-\al_{-2}\al_{-1}+{\rt{2}\o3}\al_{-1}^3\Big)
\Big({\rt{2}\o3}\til{\al}_{-3}-\til{\al}_{-2}\til{\al}_{-1}
+{\rt{2}\o3}\til{\al}_{-1}^3\Big)+\cdots\ri]|0\ket \cr
~~+&{\cal O}(\til{\la}^3).
}}
On the other hand, 
the first three terms of the discrete states with $m=j-1$ are given by
\eqn\jmonedisc{\eqalign{
|1,0,0\ket_{osc}& = \al_{-1}\til{\al}_{-1}|0\ket \cr
\lf|{3\o2},\hf,\hf\ri\ket_{osc} &= {1\o 6}(\al_{-2}-\rt{2}\al_{-1}^2)
(\til{\al}_{-2}-\rt{2}\til{\al}_{-1}^2)|0\ket \cr
|2,1,1\ket_{osc} &= {1\o4}\lf({\rt{2}\o3}\al_{-3}-\al_{-2}\al_{-1}
+{\rt{2}\o3}\al_{-1}^3\ri)
\lf({\rt{2}\o3}\til{\al}_{-3}-\til{\al}_{-2}\til{\al}_{-1}
+{\rt{2}\o3}\til{\al}_{-1}^3\ri)|0\ket 
}}
where $|j,m,m\ket_{osc}$ denotes the oscillator part of discrete state.
Comparing \Bonepri\ and \jmonedisc, we see that, for $n\leq3$, 
the order $\til{\la}^{n-1}$
term at level $n$ corresponds to the discrete state with $m=j-1$. 
We can show that this correspondence is true 
for general $n$, as follows.
Up to the choice of phase,
the discrete state operator ${\cal O}^{j,m=j-1}$ is given by
\eqn\Ojmone{\eqalign{
{\cal O}^{j,j-1}&={1\o\rt{2j}}J^{-}{\cal O}^{j,j}  \cr
&={1\o\rt{2j}}\oint{dz\o2\pi i}e^{-2iX(z)}e^{2ijX(0)}\cr
&={1\o\rt{2j}}\oint{dz\o2\pi i}z^{-2j}\exp\lf[2i(j-1)X(0)
-2i\sum_{n=1}^\infty{z^n\o n!}\del^n X(0)\ri].
}}
Using the correspondence \opstate,
the oscillator part of ${\cal O}^{j,j-1}$ is found to be
\eqn\Ojmosc{
|j,j-1\ket_{osc}={1\o\rt{2j}}\oint{dz\o2\pi i}z^{-2j}
\exp\lf[-\sum_{n=1}^\infty{\rt{2}\o n}\al_{-n}z^n\ri]|0\ket.
}
In $|B_0'\ket$ \Bextra, the order $\til{\la}^{n-1}$ term at level $n$
comes from the Young tableau $Y=(1^n)$, which has only one column. 
We denote the sum of those terms as $|b\ket$.
Using the relation 
\eqn\chionesi{
\chi_{(1^{|\si|})}(\si)=(-1)^{|\si|-\ell(\si)},
}
and recalling that $|\si,\tilde{\si}\ket$ is at the level $|\si|$,
$|b\ket$ is written as
\eqn\bvsdisc{\eqalign{
|b\ket& = -\sum_{|\si|=|\til{\si}|}(-\til{\la})^{|\si|-1}
(-1)^{|\si|-\ell(\si)+|\til{\si}|-\ell(\tilde{\si})}
{\cal N}'^{(\si,\til{\si})}|\si,\til{\si}\ket\cr
& = {1\o\til{\la}}\int_0^{2\pi}{d\th\o2\pi}\exp\lf[-\sum_{n=1}^\infty
{\rt{2}(-\til{\la})^{{n\o2}}\o n}
\lf(\al_{-n}e^{in\th}+\til{\al}_{-n}e^{-in\th}\ri)\ri]|0\ket 
-{1\o\til{\la}}|0\ket.
}}
Here, the $\th$-integral was introduced to impose the level matching
constraint $|\si|=|\tilde{\si}|$. From \Ojmosc\ and \bvsdisc, we find
\eqn\bandjmone{
|b\ket=-\sum_{j=1,{3\o2},\cdots} 2j(-\til{\la})^{2j-2}|j,j-1,j-1\ket_{osc}.
}
The coefficient $2j$ in $|b\ket$ is different from the coefficient
$\lf(\matrix{j+j-1\cr2(j-1)}\ri)=2j-1$ in \Bdisc.
This implies that the state $|j,j-1,j-1\ket$ with coefficient $\pm1$ 
is contained in the
leading term $f|D\ket$ in the decomposition \Bdecomp.
This phenomenon was observed at level 2 in \OkudaYD.

\subsec{Exponential Form of $|B\ket$}
The matrix representation of boundary state \BasgrandP\
was obtained by expanding the boundary interaction.
In this subsection, we will ``re-exponentiate'' the expression \BasgrandP.

We first consider the exponential form of $|A\ket$.
In the eigenvalue basis, the matrix representation of $|A\ket$ in \AasgrandP\  
is rewritten as
\eqn\Athetaint{
|{\cal A}\ket=\sum_{N=0}^\infty(-\til{\la})^N{1\o N!}\int\prod_{k=1}^N
{d\th_k\o2\pi}\prod_{k<l}|e^{i\th_k}-e^{i\th_l}|^2
\prod_{k=1}^N e^{2iX^{(-)}(\th_k)}|0\ket
}
where we defined $X^{(\pm)}$ by
\eqn\Xmodeexp{\eqalign{
X^{(+)}&= {i\o\rt{2}}\sum_{n>0}{1\o n}\Big[{\al_n}e^{-in\th}
+\til{\al}_ne^{in\th}\Big],\cr
X^{(-)}&= {i\o\rt{2}}\sum_{n<0}{1\o n}\Big[{\al_n}e^{-in\th}
+\til{\al}_ne^{in\th}\Big]=
-{i\o\rt{2}}\sum_{n>0}{1\o n}\Big[\al_{-n}e^{in\th}
+\til{\al}_{-n}e^{-in\th}\Big].
}}
The key observation is that 
the Vandermonde factor in \Athetaint\ can be 
written as the commutator
\eqn\Xpmcomm{\eqalign{
[iX^{(+)}(\th_k),2iX^{(-)}(\th_l)]&= -\sum_{n=1}^\infty
{1\o n}\Big[e^{-in(\th_k-\th_l)}+e^{in(\th_k-\th_l)}\Big]\cr
&= \log(1-e^{-i(\th_k-\th_l)})+\log(1-e^{i(\th_k-\th_l)}) \cr
&= \log|e^{i\th_k}-e^{i\th_l}|^2.
}}
Using the identity $e^Ae^B=e^Be^Ae^{[A,B]}$, 
we get the factor $|e^{i\th_k}-e^{i\th_l}|^2$ 
when exchanging the order of $e^{iX^{(+)}}$ and $e^{2iX^{(-)}}$:
\eqn\expXexch{
e^{iX^{(+)}(\th_k)}e^{2iX^{(-)}(\th_l)}=
e^{2iX^{(-)}(\th_l)}e^{iX^{(+)}(\th_k)}|e^{i\th_k}-e^{i\th_l}|^2.
}
To write $|A\ket$ in an exponential form, we introduce the
operator
\eqn\VopinA{
{\cal V}=\int{d\th\o2\pi}e^{2iX^{(-)}(\th)}e^{iX^{(+)}(\th)}.
}
The $N^{\rm th}$ power of ${\cal V}$ is written as
\eqn\VtoNcomp{\eqalign{
&{\cal V}^N\cr
=& \int{d\th_1\o2\pi}
e^{2iX^{(-)}(\th_1)}e^{iX^{(+)}(\th_1)}
\int{d\th_2\o2\pi}e^{2iX^{(-)}(\th_2)}e^{iX^{(+)}(\th_2)}
\cdots \int{d\th_N\o2\pi}e^{2iX^{(-)}(\th_N)}e^{iX^{(+)}(\th_N)}\cr
=& \int\prod_{k=1}^N{d\th_k\o2\pi}e^{2iX^{(-)}(\th_1)}e^{2iX^{(-)}(\th_2)}
e^{iX^{(+)}(\th_1)}|e^{i\th_1}-e^{i\th_2}|^2e^{iX^{(+)}(\th_2)}\cdots
e^{2iX^{(-)}(\th_N)}e^{iX^{(+)}(\th_N)}\cr
=& \int\prod_{k=1}^N{d\th_k\o2\pi}\prod_{k<l}|e^{i\th_k}-e^{i\th_l}|^2
\prod_{k=1}^Ne^{2iX^{(-)}(\th_k)}\prod_{l=1}^Ne^{iX^{(+)}(\th_l)}.
}}
In the second line, we changed the order of $e^{iX^{(+)}(\th_1)}$
and $e^{2iX^{(-)}(\th_2)}$. We repeated this change of ordering until all the 
$e^{iX^{(+)}}$'s are on the right of $e^{2iX^{(-)}}$'s. In this process we get
the Vandermonde factor $\prod_{k<l}|e^{i\th_k}-e^{i\th_l}|^2$.
Combining \Athetaint\ and \VtoNcomp, 
and using the fact that $X^{(+)}(\th)|0\ket=0$,
we can write $|A\ket$ in an exponential form
\eqn\Ainexpform{
| A\ket=\exp\big(-\til{\la}\,{\cal V}\big)|0\ket.
}

From \BexpA, $|B\ket$ is also written as
\eqn\Binexpform{
|B\ket=\exp\big(-\til{\la}\,{\cal V}_T\big)|N\ket,
}
where ${\cal V}_T$ is the tachyon operator given by
\eqn\VTop{
{\cal V}_T=\exp\lf(-\sum_{n=1}^\infty{1\o n}\al_{-n}\til{\al}_{-n}\ri)
{\cal V}\exp\lf(\sum_{n=1}^\infty{1\o n}\al_{-n}\til{\al}_{-n}\ri)
=\int{d\th\o2\pi}e^{2iX^{(-)}}e^{iX^{(+)}-iX^{(-)}}. 
}
Note that the ordering in ${\cal V}_T$ is different from
the ordinary normal ordering.
If we define the ``bare'' coupling $\la_0$ as the coefficient in front of the
normal ordered operator
\eqn\lazero{
\la_0 e^{iX^{(-)}}e^{iX^{(+)}}=\la e^{2iX^{(-)}}
e^{iX^{(+)}-iX^{(-)}},
}
then $\la$ and $\la_0$ are related by
\eqn\lalazero{
\la=\la_0 e^{\hf[iX^{(+)},iX^{(-)}]}=\la_0|e^{i\th}-e^{i\th}|^{\hf}.
}
To make the coupling $\la$ finite, we should introduce a cut-off and
take the limit $\la_0\riya\infty$.
One might worry that this renormalization would lead to the breaking
of the conformal invariance.
In the next subsection, we directly check the conformal invariance of 
the matrix representation of $|B\ket$ \BasgrandP.

\subsec{Conformal Invariance}
Let us check the conformal invariance of the boundary state $|B_{U(N)}\ket$
with fixed $N$ 
\eqn\BUgeneral{
|B_{U(N)}\ket\equiv \int_{U(N)}dU|B_U\ket,\quad 
(\al_n+\til{\al}_{-n})|B_U\ket=\ga\Tr\,U^n|B_U\ket.
}
Here we slightly generalized the condition \BUbc\ by replacing the coefficient 
$\rt{2}$ in front of $\Tr\,U^n$ by a constant $\ga$.
The level matching condition $(L_0-\til{L}_0)|B_{U(N)}\ket=0$ 
follows from the $U(1)$ symmetry $U\riya e^{i\al}U$
of the matrix integral. 
To see the invariance under $L_n-\til{L}_{-n}$,
it is convenient to write  
the measure $dU$
as a contour integral around $z=0$ 
\eqn\Umeasure{
\int_{U(N)} dU={1\o N!}\oint\prod_{k=1}^N{dz_k\o2\pi iz_k}\prod_{i<j}(z_i-z_j)
(z_i^{-1}-z_j^{-1}).
}
Then the action of $L_n-\til{L}_{-n}$ is represented
by the change of integration variables
\eqn\zkshift{
z_k\riya z_k+\vep z_k^{n+1}.
}
By writing explicitly the variation of $|B_{U(N)}\ket$ under \zkshift,
we can show that the condition $(L_n-\til{L}_{-n})|B_{U(N)}\ket=0$ 
holds if and only if $\ga^2=2$ \foot{
In the literature of ${\rm YM}_2$, the correspondence between the oscillators
and the trace of unitary matrix
was usually written as $\al_n+\til{\al}_{-n}\lrya\Tr\,U^n$, {\it i.e.},
$\ga=1$ \refs{\DouglasWY,\CordesFC}.
}.
The two allowed values $\ga=\pm\rt{2}$ correspond to the tachyon configurations
$T=\la e^{\pm X^0}$. After taking into account the open string creation and
the closed string emission, the conformal invariance will be broken due to
the divergence in both channels \refs{\LLM,\KarczmarekXM}.

\subsec{The Leading Term $|B_f\ket=f|D\ket$}
The leading term of $|B\ket$ in the decomposition \Bdecomp\ has the form
\eqn\fDterm{
|B_f\ket\equiv f|D\ket=\int dx^0 \,f(x^0)|D_{x^0}\ket.
}
Therefore, $f(x^0)$ can be thought of as the distribution function of
SD-branes localized on the real time axis.
$|B_f\ket$ roughly corresponds to
the Lorentzian FZZ state \BFZZ
\eqn\BfFZZ{
|B_f\ket \simeq |B_{\rm LFZZ}\ket.
}
The wave function of the bulk primary operator $e^{iEX^0}$ seems to agree on
both sides. However, the oscillator structure might be different between 
$|B_f\ket$ and $|B_{\rm LFZZ}\ket$.
In computing the wave function of $e^{iEX^0}$, {\it i.e.}, the Fourier 
transform of $f(x^0)$ \fxzero, we should be careful about the possible 
divergence coming from the region $x^0\sim-\infty$
\eqn\fintEx{
i\h{f}(E)=i\int_{-\infty}^\infty dx^0e^{iEx^0}f(x^0)
=i\int_{-\infty}^\infty dx^0{e^{iEx^0}\o1+\la e^{x^0}}.
}
In order to make this integral convergent, 
we should add an imaginary part to $E$ as $E\riya E-i\vep$, where $\vep>0$.
Then the integral becomes well-defined and
the Fourier transform of $f$ is computed as
\eqn\FtrfEvep{
i\h{f}(E)=\la^{-iE}{\pi\o\sinh\pi (E-i\vep)}.
}
By splitting $\h{f}(E)$ into the principal value part and the discontinuity
part,  $\h{f}(E)$ is written as
\eqn\Ftrfoff{
i\h{f}(E)=\pi i\cob(E)+ \la^{-iE}
{\bf P}\lf({\pi\o\sinh\pi E}\ri),
}
where ${\bf P}$ stands for the principal value
\eqn\Pval{
{\bf P}F(E)=\lim_{\vep\riya0}{ F(E+i\vep)+ F(E-i\vep)\o2}.
}
Since $1/\sinh\pi E$ is an odd function of $E$, 
there is no singularity at $E=0$ in the principal value part
\eqn\Psinhzero{
\lim_{E\riya0}{\bf P}\lf({\pi\o\sinh\pi E}\ri)=0,
}
thus
the singularity of $\h{f}(E)$
at $E=0$ is contained only in the delta-function term in \Ftrfoff.
Since the open string tachyon stays near the top of the potential
when $x^0<-\log\la$, the unstable brane can support the ends of
open string for a very long time before it decays.
This can be thought of as the origin of the singular term
$\cob(E)$ in $\h{f}(E)$ \McGreevyEP.

The same form of the Fourier transform can be obtained by writing $f$ as 
\eqn\fxzerodef{\eqalign{
f(x^0) & = \hf-\hf\tanh\lf({x^0+\log \la\o2}\ri) \cr
& = \hf-\sum_{n=-\infty}^\infty{1\o x^0+\log \la-\pi i(2n+1)}.
}}
The constant term $\hf$ gives to the delta-function in \Ftrfoff.
Note that the pole $x^0=-\log \la+\pi i(2n+1)$ corresponds to the position of
the imaginary branes discussed in \refs{\LLM,\GaiottoRM}.
We can also write $f(x^0)$ as the resolvent of the
matrix coordinate ${\bf X}^0$
of D-instantons 
\eqn\fasresolv{
f(x^0)=\hf-\Tr{1\o x^0-{\bf X}^0}
}
where ${\bf X}^0={\rm diag}(-\log \la+\pi i(2n+1))_{n\in\Z}$.
Similarly, in the superstring case, 
the source of RR-field generated by the 
rolling tachyon can be written as a super-trace version of resolvent.
 
When computing the Fourier transform of $f$, we close the contour 
of $x^0$-integral in the upper (lower) 
half plane if $E>0$ $(E<0)$, and pick up the poles of $f(x^0)$. 
Therefore, the in (out) state corresponds to
the semi-infinite array of D-instantons sitting
along the line $x^0=-\log\la$
on the upper (lower) half plane.
In this way our decomposition \Bdecomp\ naturally explains the
appearance of D-instanton array in the imaginary part of 
annulus amplitude \LLM. The computation in \LLM\ amounts to 
ignore the term $|B_0'\ket$. One may argue that
$|B_0'\ket$ does not contribute to the imaginary part 
of annulus amplitude since all terms in $|B_0'\ket$ 
are``off-shell''.

\newsec{Annulus Amplitude and Unitary Two-Matrix Integral}
In this section, we consider the annulus amplitude using the matrix
representation of boundary state \BasgrandP.
Since the annulus has two boundary circles and the boundaries of worldsheet
act as the space of eigenvalues of unitary matrix,
the annulus amplitude is written as a unitary two-matrix integral.

Before writing the matrix representation of annulus,
we emphasize that in general the Lorentzian annulus amplitude
is not directly related to the Euclidean one.
In the Euclidean case, the Ishibashi state is normalized as
\eqn\jmIshi{
\bra\bra j,m,\til{m}|q^{\hf(L_0+\til{L}_0)-{1\o24}}|j',m',\til{m}'\ket\ket
=\cob_{j,j'}\cob_{m,m'}\cob_{\tilde{m},\tilde{m}'}
{q^{j^2}-q^{(j+1)^2}\o\eta(q)}.
}
The conservation of momentum $\cob_{m,m'}$ is due to the 
zero-mode integration.
In the Lorentzian case, we cannot integrate over $x^0$
term by term, since each term has the form $e^{mx^0}$ which diverges 
at late times.
As we mentioned in section 2.1, in the timelike Liouville theory it is useful to
integrate out the non-zero modes first. 
In this paper, we will try to define the annulus amplitude
by first contracting the oscillator part 
and then performing a resummation to make the integrand Fourier
transformable, if possible. 

So we consider the annulus amplitude {\it without} the zero-mode integration:
\eqn\AnnulusBB{
Z(\til{\la}_1,\til{\la}_2)
=\langle B(\til{\la}_1)|q^{{1\over 2}(N+\tilde{N})-{1\over 24}}
|B(\til{\la}_2)\rangle.
}
As before,
we treat the zero-mode as a part of the coupling $\til{\la}=\la e^{x^0}$.
To compute $Z(\til{\la}_1,\til{\la}_2)$, we can make use of the formula
of oscillator contraction
\eqn\oscformula{\eqalign{
&\bra 0|\exp\Big[{a_iT_{ij}b_j+\la_i^a a_i+\la^b_i b_i}\Big]
\exp\Big[{a_i^\dag S_{ij}b_j^\dag+\mu_i^a a_i^\dag
+\mu^b_i b_i^\dag}\Big]|0\ket\cr
=~&{1\o\det_{ij}(1-TS)}\exp\lf[\la^a_i\lf({1\o1-TS}\ri)_{ij}\mu^a_j
+\mu^b_i\lf({1\o1-TS}\ri)_{ij}\la^b_j\ri.\cr
&\hskip27mm\lf.+\mu^b_i\lf({T\o1-TS}\ri)_{ij}\mu^a_j
+\la^a_i\lf({S\o1-TS}\ri)_{ij}\la^b_j\ri],
}}
where $\la^{a,b}_i,\mu^{a,b}_i,S_{ij}$ and $T_{ij}$ are c-number parameters,
and $a_i, b_i$ are normalized oscillators
\eqn\abcomm{
[a_i,a^\dag_j]=[b_i,b_j^\dag]=\cob_{ij},\quad[a_i,b^\dag_j]=0.
}
This formula assumes that
$S_{ij}$ and $T_{ij}$ are symmetric matrices and commute with each other.
We use this formula by identifying
$a_i\lrya \al_n/\rt{n}$ and $b_i\lrya \til{\al}_n/\rt{n}$.
Then $Z(\til{\la}_1,\til{\la}_2)$ is evaluated as
\eqn\ZBBasUint{
Z(\til{\la}_1,\til{\la}_2)= {1\over \eta(q)}
\sum_{N,M=0}^\infty(-\til{\la}_1)^N(-\til{\la}_2)^M Z_{N,M},
}
where $Z_{N,M}$ is given by
\eqn\ZNMdef{\eqalign{
Z_{N,M}&=\int_{U(N)\times U(M)}\hskip-12mm dUdV~~
\exp\left[\sum_{n=1}^\infty{2\over n(1-q^n)}
\Big(q^{{n\over2}}{\rm Tr}\,U^n{\rm Tr}\,V^{-n}
+q^{{n\over2}}{\rm Tr}\,U^{-n}{\rm Tr}\,V^{n}\right. \cr
& \hskip 52mm \left.
-q^{n}{\rm Tr}\,U^n{\rm Tr}\,U^{-n}
-q^{n}{\rm Tr}\,V^n{\rm Tr}\,V^{-n}\Big)\right].
}}
Namely, the annulus amplitude is written as the grand canonical
partition function of
the unitary two-matrix model with the double trace interactions.

$Z_{N,M}$ can be also written in a determinant form
by expanding the factor  $(1-q^n)^{-1}
=\sum_{m\geq0}q^{nm}$, and summing over $n$ using 
$\sum_{n>0}x^n/n=-\log(1-x)$. In this way, we get
\eqn\ZNMdetform{\eqalign{
Z_{N,M}&=\int_{U(N)\times U(M)}\hskip-12mm dUdV~~
\exp\left[\sum_{n,m=1}^\infty{2\over n}
q^{{n(m-\hf)}}\Big\{{\rm Tr}(U\tens V^{-1})^n
+{\rm Tr}(U^{-1}\tens V)^{n}\Big\}\ri.\cr
& \hskip 44mm \lf.-{2\o n}q^{nm}\Big\{{\rm Tr}(U\tens U^{-1})^n+
{\rm Tr}(V\tens V^{-1})^n\Big\}\right] \cr
&= \int_{U(N)\times U(M)}\hskip-12mm dUdV~~
\prod_{m=1}^\infty{\det(1-q^mU\tens U^{-1})^2\det(1-q^mV\tens V^{-1})^2
\o\det(1-q^{m-\hf}U\tens V^{-1})^2\det(1-q^{m-\hf}U^{-1}\tens V)^2}.
}}
Here we have also used the relation $\Tr A\Tr B=\Tr(A\tens B)$.

The large $N,M$ limit of $Z_{N,M}$ is obtained by the Gaussian integral \largeN
\eqn\LargeNZMN{
\lim_{N,M\riya\infty}Z_{N,M}
=\prod_{n=1}^\infty\det\lf[\lf(\matrix{1&0\cr0&1}\ri)
-\lf(\matrix{{-2q^n\o1-q^n}&{2q^{n\o2}\o1-q^n}
\cr{2q^{n\o2}\o1-q^n}&{-2q^n\o1-q^n}}\ri)\ri]^{-1}
=1.
}
Therefore, 
if we replace all the finite $N$ integrals in \ZBBasUint\ by 
the large $N$ value 
\LargeNZMN, $Z(\til{\la}_1,\til{\la}_2)$ reduces to the annulus amplitude 
between the leading term $|B_f\ket$
\eqn\ZBBlargeN{
Z(\til{\la}_1,\til{\la}_2)\sim {1\o\eta(q)}\sum_{N,M=0}^\infty
(-\til{\la}_1)^N(-\til{\la}_2)^M=
{1\o\eta(q)}{1\o(1+\til{\la}_1)(1+\til{\la}_2)}=\bra B_{f_1}|
q^{\hf(N+\til{N})-{1\o24}}|B_{f_2}\ket.
} 
From \ZNMdetform\ and \LargeNZMN, we can see that 
the large $N$ limit of $Z_{N,M}$ 
is equivalent to the limit $\lim_{q\riya0}Z_{N,M}=1$.
Namely, the large $N$ limit amounts to neglect the contribution from
the exponentially growing timelike oscillators found in \OkudaYD.
Note that the partition function of $bc$-ghosts $\eta(q)^2$
does not cancel the $q$-dependence in $Z_{N,M}$.

The amplitude with zero-mode integrated is given by
\eqn\Zwithzeroint{
\bra B|q^{\hf(L_0+\til{L}_0)-{1\o24}}|B\ket
=\int_{-\infty}^\infty {dE\o2\pi}
dx^0dy^0
~q^{-{1\over4}E^2}e^{-iE(x^0-y^0)}Z(\la e^{x^0},\la e^{y^0}).
}
Without some appropriate prescription,
all three integrals $\int dE,\int dx^0$ and $\int dy^0$ are ill-defined
because the integrand is divergent at infinity. 
Our proposal is to perform a resummation in \ZBBasUint\ 
in order to make the integral 
over the zero-modes $x^0$ and $y^0$ convergent.
If this resummation is possible and the Fourier transform of 
$Z(\la e^{x^0},\la e^{y^0})$ becomes well-defined, 
the remaining integral $\int dE$ can be treated in the same 
way as in \KarczmarekXM.

\subsec{Some Examples of Matrix Integral}
In this subsection, we compute the integral $Z_{N,M}$ \ZNMdef\ for some cases.
Since $Z_{N,M}$ is symmetric in $N$ and $M$,
$Z_{N,M}=Z_{M,N}$, it is sufficient to consider
the cases $N\geq M$.
Let us first compute the integral $Z_{N,M}$ with
$M=0$, 
\eqn\UNintex{\eqalign{
Z_{N,0}&= \int_{U(N)}dU\exp\lf(-\sum_{n=1}^\infty{2q^n\o n(1-q^n)}
\Tr\,U^n\Tr\,U^{-n}\ri) \cr
&=\int_{U(N)}dU\prod_{n=1}^\infty\det(1-q^nU\tens U^{-1})^2.
}}
This integral also appears in the amplitude 
$\bra N|q^{\hf(N+\til{N})-{1\o24}}|B\ket$
between the Neumann state and the rolling tachyon state. 
Since the $U(1)$ part of $U(N)$ integral is decoupled in \UNintex,
$\int_{U(N)}$ can be replaced by $\int_{SU(N)}$.
This integral is computable thanks to the fact 
that the square root of the integrand
is proportional to the Weyl denominator of affine $A_{N-1}$\foot{
Similar integral has appeared in the inner product of 
the wave-functions of Chern-Simons theory on $T^2\times I$ \DouglasEX
}.
Using the Weyl denominator
formula of $\h{A}_{N-1}$ \Kac, we can expand the square root of integrand
in terms of the $SU(N)$ characters $\chi_{\la}(U)$
\eqn\chexpdet{\eqalign{
\prod_{m=1}^\infty\det(1-q^mU\tens U^{-1})&=\prod_{n=1}^\infty(1-q^n)
\sum_{\al\in\La_R}\chi_{N\al}(U)
q^{{1\o2N}C_2(N\al)}\cr
&=\prod_{n=1}^\infty(1-q^n)
\sum_{\al\in\La_R}\chi_{N\al}(U^{-1})
q^{{1\o2N}C_2(N\al)},
}}
where $\La_{R}$ is the root lattice of $SU(N)$, and 
$C_2(\la)=(\la,\la+2\rho)$
is the second Casimir with $\rho$ being
the half of the sum of positive roots.\foot{
Note that 
eq.\chexpdet\ is reminiscent of the disk amplitude of YM$_{2}$ with finite
area. Here the closed string proper time plays the role of the area.
However this analogy is not exact, since in \chexpdet\ only the sublattice
$N\La_R$ of the weight lattice $\La_w$ contributes to the summation.
} 
Using the orthogonality of the characters, 
$Z_{N,0}$ becomes
\eqn\ZUNasTheta{
Z_{N,0}=\prod_{n=1}^\infty(1-q^n)^2\sum_{\al\in\La_R}q^{{1\o N}C_2(N\al)}.
}
The summation over $\La_R$
can be performed
by substituting the following ``ground state'' configuration 
$U_0$ into \chexpdet\  \refs{\Macdonald,\Jinzenji}
\eqn\masterU{
U_0={\rm diag}(\om^{n_F},
\om^{n_F-1},\cdots, \om^{-n_F});\qquad \om=e^{2\pi i/N},~~n_F={N-1\o2}.
}
Note that the eigenvalues of $U_0$ are distributed
on the unit circle in a $\Z_N$-symmetric way.
The special property of $U_0$ is that $\chi_{N\al}(U_0)=1$ 
for all $\al\in\La_R$.
Therefore, after replacing $q$ by $q^2$ in \chexpdet, the summation in 
\ZUNasTheta\ is written as\foot{This is a version of the Macdonald identity
\Macdonald. In the combinatorial language, this sum
is related to the generating function of the ``$N$-core'' of partitions 
\Macbook.}
\eqn\Macid{
\sum_{\al\in\La_R}
q^{{1\o N}C_2(N\al)}
=\prod_{m=1}^\infty{\det(1-q^{2m}U_0\tens U_0^{-1})
\o1-q^{2m}}=\prod_{m=1}^\infty{(1-q^{2Nm})^N
\o1-q^{2m}}.
} 
One may think of $U_0$ as a kind of ``master field''. 
In the large $N$ limit, the eigenvalue distribution of $U_0$ approaches 
the uniform distribution on the unit circle.
Finally, $Z_{N,0}$ is found to be
\eqn\ZUNeta{
  Z_{N,0}=
 \prod_{m=1}^\infty{(1-q^m)(1-q^{2Nm})^N\o 1+q^m}.
}

In the general case $N,M\not=0$, the square root of the integrand of $Z_{N,M}$
is proportional to the Weyl denominator of affine superalgebra
$\h{A}(N-1,M-1)$ \Wakimoto\foot{
In \Wakimoto, the Weyl denominator which is relevant for our integral
is denoted as $\h{R}_T$.
}.
Although the denominator identity of $\h{A}(N-1,M-1)$
is known unless $N=M$, it is not in a form of the character expansion, which 
was useful for the $U(N)$ case \chexpdet.
Because of this, so far we have not succeeded
in computing the integral $Z_{N,M}$ for general $N,M$.
Below, we compute a few terms of $Z_{N,M}$. 

For example,
using the $\h{A}(1,0)$ denominator identity \refs{\Wakimoto,\MatsuoCJ}
\eqn\idtwoone{\eqalign{
\h{R}(x,y)=&\prod_{n=1}^\infty{(1-q^n)^2(1-q^nxy)(1-q^{n-1}x^{-1}y^{-1})\o
(1-q^{n-\hf}x)(1-q^{n-\hf}x^{-1})(1-q^{n-\hf}y)(1-q^{n-\hf}y^{-1})}\cr
=&\lf(\sum_{m,n\geq0}-\sum_{m,n<0}\ri)x^my^nq^{mn+\hf(m+n)},
}}
$Z_{2,1}$ is evaluated as
\eqn\Ztwooneform{\eqalign{
Z_{2,1}&=\prod_{m=1}^\infty(1-q^m)^2{1\o2!}\int_0^{2\pi}
{d\th_1d\th_2d\phi\o(2\pi)^3}
\h{R}(e^{i\th_1-i\phi},e^{-i\th_2+i\phi})
\h{R}(e^{-i\th_1+i\phi},e^{i\th_2-i\phi})\cr
&=\prod_{m=1}^\infty(1-q^m)^2\sum_{n=0}^\infty{q^n\o1-q^{2n+1}}.
}}
We can also compute $Z_{1,1}$ defined by
\eqn\Zoneonedef{
Z_{1,1}=\int_0^{2\pi}{d\th\o2\pi}\int_0^{2\pi}{d\phi\o2\pi}
\prod_{m=1}^\infty{(1-q^m)^4\o (1-q^{m-\hf}e^{i\th-i\phi})^2
(1-q^{m-\hf}e^{-i\th+i\phi})^2}.
}
This integral can be evaluated by using the identity,
\eqn\idoneone{
\prod_{n=1}^\infty{(1-q^n)^4\o(1-q^{n-\hf}z)^2(1-q^{n-\hf}z^{-1})^2}=
\sum_{m,n\geq0}^\infty(m+n+1)z^{m-n}q^{mn+\hf(m+n)},
}
which is obtained as a limit of \idtwoone,
and we get
\eqn\Zoneoneform{
Z_{1,1}=\sum_{n=0}^\infty(2n+1)q^{n(n+1)}.
}

Since we do not have a closed form of $Z_{N,M}$ for general $N,M$,
\foot{
We have computed first few terms of $Z_{3,1}$ and $Z_{2,2}$
by using {\tt Mathematica}: 
\eqn\ZMathe{\eqalign{
Z_{3,1}&=1 - 3 q^2 - q^4 + 6 q^6 + 7 q^8 - 10 q^{10} - 9 q^{12} - q^{14}
+{\cal O}(q^{17}),\cr
Z_{2,2}&=1+8q^3+27q^8+64q^{15}+{\cal O}(q^{17}).
}}
}
we cannot discuss whether or not the resummation in \ZBBasUint\
is possible. In the next section, we consider an example of annulus amplitude
in which we can perform a resummation explicitly.

\newsec{Annulus Amplitude between $|B_f\ket$ and $|B\ket$}
In this section, we show that if we replace one of the boundary states
in \AnnulusBB\ by its leading term 
the contribution from the exponentially growing timelike
oscillators \OkudaYD\ can be resummed.
Namely, we consider the following amplitude without the zero-mode integral: 
\eqn\ZBfB{
\bra B_f|q^{\hf(N+\til{N})-{1\o24}}|B\ket
=f\bra D|q^{\hf(N+\til{N})-{1\o24}}|B\ket.
}
This can be thought of as a part of the amplitude between
two $|B\ket$'s.

\subsec{Matrix Integral}
Since the zero-mode dependence of $\bra B_f|$ is factorized as $f\bra D|$,
we focus on the oscillator contraction between
$\bra D|$ and $|B\ket$ :
\eqn\ZDBdef{
Z_{DB} = \bra D|q^{\hf(N+\til{N})-{1\o24}}|B\ket
= q^{-{1\o24}}\prod_{n=1}^\infty(1+q^n)^{-1}
\sum_{N=0}^\infty(-\til{\la})^N F_N,
}
where $F_N$ is given by
\eqn\FNqQm{\eqalign{
F_N&=\int_{U(N)}dU\exp\lf(\sum_{n=1}^\infty
{2q^n\o n(1+q^n)}\Tr\,U^n\Tr\,U^{-n}\ri) \cr
&=\int_{U(N)}dU
\prod_{n=1}^\infty
{\det(1-q^{2n}U\tens U^{-1})^2\o\det(1-q^{2n-1}U\tens U^{-1})^2}.
}}
$F_0$ and $F_1$
are easily found to be
\eqn\Fzeroone{
F_0=1,\quad F_1=\prod_{n=1}^\infty 
{(1-q^{2n})^2\o(1-q^{2n-1})^2}.
}
As before, 
the square root of the integrand of
$F_N$ is related to the Weyl denominator of some affine superalgebra.
In this case, the relevant algebra is the so-called 
$Q(N-1)$ superalgebra \Wakimoto.
For example,
using the denominator identity of $Q(1)$ affine superalgebra
\eqn\Rofx{
\h{R}(x)=\prod_{n=1}^\infty{(1-q^{2n})^2(1-q^{2n-2}x)(1-q^{2n}x^{-1})\o
(1-q^{2n-1})^2(1-q^{2n-1}x)(1-q^{2n-1}x^{-1})}=
\sum_{k=-\infty}^\infty{q^k\o1-q^{2k+1}x},
}
$F_2$ is evaluated as
\eqn\Ftwo{
F_2={1\o2!}\int_{0}^{2\pi}{d\th_1d\th_2\o(2\pi)^2}
\h{R}(e^{i\th_1-i\th_2})\h{R}(e^{-i\th_1+i\th_2})
=\sum_{k=0}^\infty{q^{2k}\o(1-q^{2k+1})^2}.
}
Similarly, $F_3$ is obtained by using the denominator identity of $Q(2)$ 
\refs{\Wakimoto,\Zagier}
\eqn\Fthree{
F_3=\prod_{n=1}^\infty{(1-q^{2n})^2\o(1-q^{2n-1})^2}
\sum_{k=0}^\infty{q^{2k}\o(1-q^{2k+2})^2}.
}

For $Q(N-1)$ affine superalgebra with $N>3$,
there is only a conjectured denominator formula which has not 
been proved yet \Wakimoto.
Instead of computing the matrix integral directly,
in the next subsection we consider a different approach.

\subsec{Computation of $Z_{DB}$}
$Z_{DB}$ can be obtained 
by making use of the knowledge of Euclidean annulus amplitude. 
As we  emphasized in section 3, in general the Lorentzian annulus amplitude
is not directly related to the Euclidean one.
In particular, we need the matrix element {\it without} the
zero-mode integration
\eqn\jmIshiosc{
{}_{osc}\bra\bra j,m,\til{m}|q^{\hf(N+\til{N})-{1\o24}}
|j',m',\til{m}'\ket\ket_{osc}.
}
In general, this amplitude is not diagonal in $m$.\foot{
From \zeroalasN, the off-diagonal metrix elements \jmIshiosc\
correspond to the integrals $Z_{N,M}$ with $N\not=M$.
Especially, these terms are necessary to have the factorized form $f_1f_2$
in \ZBBlargeN.
}
However, $Z_{DB}$ is related to the Euclidean amplitude as follows.
The amplitude $Z_{DB}$ without zero-mode integral is written as
\eqn\IshioscZDB{
Z_{DB}=\bra D|q^{\hf(N+\til{N})-{1\o24}}|B\ket
=\bra D|q^{\hf(N+\til{N})-{1\o24}}\lf[\sum_{j,m\geq0}
\lf(\matrix{j+m\cr2m}\ri)(-1)^m\til{\la}^{2m}|j,m,m\ket\ket_{osc}\ri].
}
The important fact here is that the matrix element
\eqn\DqIshiosc{
\bra D|q^{\hf(N+\til{N})-{1\o24}}
|j,m,m\ket\ket_{osc}
}
appearing in \IshioscZDB\
is equal to the Euclidean amplitude with zero-mode integrated
\eqn\DqIshiEuc{
q^{-m^2}\bra D;m,m|q^{\hf(L_0+\til{L}_0)-{1\o24}}|j,m,m\ket\ket,
}
where we defined
\eqn\Dmmdef{
\bra D;m,m|=\Big\bra \al_0=\til{\al}_0=\rt{2}m\Big|
\exp\lf(\sum_{n=1}^\infty{1\o n}\al_n\til{\al}_n\ri).
}
The factor $q^{-m^2}$ in \DqIshiEuc\ cancels the weight 
from the zero-mode $\hf\al_0^2=m^2$, thus \DqIshiEuc\ is equal to the 
oscillator contraction \DqIshiosc. In other words, we add the Euclidean 
zero-mode by hand and subtract its effect by the factor $q^{-m^2}$.
Again, we emphasize that we use 
the Euclidean amplitude \DqIshiEuc\ just as a technical tool to compute the
Lorentzian amplitude.
Now $Z_{DB}$ is written as
\eqn\ZDBandEuc{\eqalign{
Z_{DB}
&=\sum_{j,m\geq0}
\lf(\matrix{j+m\cr2m}\ri)(-1)^m\til{\la}^{2m}q^{-m^2}
\bra D;m,m|q^{\hf(L_0+\til{L}_0)-{1\o24}}|j,m,m\ket\ket \cr
&=\sum_{m'\in\hf\Z}\bra D;m',m'|q^{\hf(L_0+\til{L}_0)-{1\o24}}
\lf[\sum_{j,m\geq0}q^{-m^2}
\lf(\matrix{j+m\cr2m}\ri)(-1)^m\til{\la}^{2m}|j,m,m\ket\ket\ri] \cr
&=\bra D_{SU(2)}|q^{\hf(L_0+\til{L}_0)-{1\o24}}\lf[\sum_{j,m\geq0}q^{-m^2}
\lf(\matrix{j+m\cr2m}\ri)(-1)^m\til{\la}^{2m}|j,m,m\ket\ket\ri] 
}}
where $\bra D_{SU(2)}|$ is given by \CallanUB
\eqn\DSUtwo{
\bra D_{SU(2)}|=\sum_{m'\in\hf\Z}\bra D;m',m'|=\sum_{j,m}\bra\bra j,m,m|(-1)^j.
}
In the second line of \ZDBandEuc, we used the momentum conservation
in the Euclidean amplitude.
Using \jmIshi, we finally find
\eqn\ZDBconech{
Z_{DB}= {1\o\eta(q)}\sum_{j,m\geq0}\bigg(\matrix{j+m\cr 2m}\bigg)
(-1)^{j+m}\til{\la}^{2m}q^{-m^2}\Big(q^{j^2}-q^{(j+1)^2}\Big).
}
Note that we cannot use this trick to compute 
$\bra B|q^{\hf(N+\til{N})-{1\o24}}|B\ket$ since in this case
we need to know the general
``off-diagonal'' amplitude \jmIshiosc.\foot{
The diagonal integral $Z_{N,N}$ \ZNMdef\
can be obtained by using this trick:
\eqn\ZNNIshi{
Z_{N,N}=\sum_{k=0}^\infty\lf[\lf(\matrix{N+k\cr N}\ri)^2
-\lf(\matrix{N-1+k\cr N}\ri)^2\ri]q^{k(k+N)}.
}
For $N=1,2$, eq.\ZNNIshi\ reproduces the direct matrix computation
\Zoneoneform\ and \ZMathe.
} 
One can easily check that the first two terms of \ZDBconech\
in the $\til{\la}$-expansion 
agree with the matrix computation \Fzeroone.
The agreement 
at the order $\til{\la}^2$ and $\til{\la}^3$ requires the following identities:
\eqna\Fprodid
$$\eqalignno{
F_2=&\sum_{k=0}^\infty{q^{2k}\o(1-q^{2k+1})^2}
=\prod_{n=1}^\infty{1+q^n\o 1-q^n}
\sum_{k=0}^\infty(-1)^k(k+1)^2q^{k(k+2)},  &\Fprodid a\cr
F_3=&\prod_{n=1}^\infty{(1-q^{2n})^2\o(1-q^{2n-1})^2}
\sum_{k=0}^\infty{q^{2k}\o(1-q^{2k+2})^2} \cr
=&\prod_{n=1}^\infty{1+q^n\o1-q^n}\sum_{k=0}^\infty{(-1)^k\o6}
(k+1)(k+2)(2k+3)q^{k(k+3)}. &\Fprodid b
}$$ 
Eqs.\Fprodid{a,b} 
can be proved by taking a derivative of the Jacobi's triple product identity 
(see appendix A for details).
We believe that the two expressions \ZDBdef\ and \ZDBconech\ are equivalent
for all orders in $\til{\la}$.

$Z_{DB}$ can also be written in the form of $SU(2)_{k=1}$ character
as in \Sugawara
\eqn\ZDBchexp{
Z_{DB} = {q^{-\qu\del_0^2}\o\eta(q)}
\lf[1+2\sum_{n=1}^\infty(-1)^nq^{n^2}\cos 2n\al
-2\sum_{n=0}^\infty(-1)^nq^{(n+\hf)^2}\sin(2n+1)\al\ri],
}
where we defined
\eqn\phaseinZDB{
\al=\arcsin\Big(\hf\til{\la}\Big).
}
One might think that $\til{\la}=2$ is a critical value above which
$\al$ aquires an imaginary part and the amplitude becomes divergent 
as $\til{\la}\riya\infty$.
But this is merely a rephrasing of the fact that at a fixed
level the coupling to the timelike oscillators grows exponentially.
However, there is a room to
perform a resummation over terms at different levels.
In the next subsection, we will consider this possibility.

\subsec{Resummation of Timelike Oscillators}
To perform a resummation in \ZDBconech, 
we write $j=m+k$ $(k=0,1,\cdots)$ and sum over $m$ with fixed $k$,
and then sum over $k$.
The amplitude with fixed $k=j-m$ is
\eqn\Zfixedk{\eqalign{
Z^{(k)}_{DB}&={1\o\eta(q)}\sum_{m=0,\hf,1,\cdots}^\infty(-1)^{2m+k}
\lf(\matrix{2m+k\cr 2m}\ri)
\Big[(\til{\la}q^k)^{2m}q^{k^2}-(\til{\la}q^{k+1})^{2m}q^{(k+1)^2}\Big]\cr
&={(-1)^k\o\eta(q)}\sum_{n=0}^\infty
\lf(\matrix{n+k\cr n}\ri)
\Big[(-\til{\la}q^k)^{n}q^{k^2}-(-\til{\la}q^{k+1})^{n}q^{(k+1)^2}\Big] \cr
&={(-1)^k\o\eta(q)}\lf[{q^{k^2}\o(1+\til{\la}q^k)^{k+1}}
-{q^{(k+1)^2}\o(1+\til{\la}q^{k+1})^{k+1}}\ri].
}}
Therefore we get
\eqn\ZDBresum{\eqalign{
Z_{DB}
&={1\o\eta(q)}\sum_{k=0}^\infty(-1)^k\lf[{q^{k^2}\o(1+\til{\la}q^k)^{k+1}}
-{q^{(k+1)^2}\o(1+\til{\la}q^{k+1})^{k+1}}\ri]\cr
 &= {1\o\eta(q)}\lf[{1\o1+\til{\la}}+\sum_{k=1}^\infty(-1)^kq^{k^2}
{2+\til{\la}q^k\o(1+\til{\la}q^k)^{k+1}}\ri].
}}
This expression is finite in the limit $\til{\la}\riya\infty$.
The first term in the second line is 
$f\bra D|q^{\hf(N+\til{N})-{1\o24}}|D\ket$ which
comes from the $k=0$ term, {\it i.e.}, the
$m=j$ term studied in the original work by Sen.
The second term is $\bra D|q^{\hf(N+\til{N})-{1\o24}}|B'_0\ket$,
which includes the contribution of the exponentially growing terms found 
in \OkudaYD. Namely, after performing a resummation over different levels,
the contribution from $|B_0'\ket$ becomes finite! 

Now we can perform the Fourier transformation of $Z_{DB}$ safely.
Note that
\ZDBresum\ exhibits an interesting interplay between the zero-mode $x^0$
and the closed string proper time $t$ in $q=e^{-4t}$. 
The Fourier transform of $Z_{DB}$ is given by
\eqn\FtrfZDB{
i\int dx^0 e^{iEx^0}q^{-\qu E^2}Z_{DB}
={\pi\la^{-iE}\o\sinh\pi E}\sum_{k=0}^\infty {(iE)^k\o k!}
\Big[\chi_{E+2ik}(q)-\chi_{E+2i(k+1)}(q)\Big],
}
where $\chi_E(q)$ is the character in the closed string channel
\eqn\chiE{
\chi_{E}(q)={1\o\eta(q)}q^{-\qu E^2}.
}
For simplicity, we ignored the delta-fucntion $\cob(E)$ 
discussed in section 2.5.
After all, the annulus amplitude between $\bra B_f|$ and $|B\ket$
is given by
\eqn\ZBfBinclosed{
\bra B_f|q^{\hf(L_0+\til{L}_0)-{1\o24}}|B\ket
=\int{dE\o2\pi}{\pi^2\o\sinh^2\pi E}\sum_{k=0}^\infty {(iE)^k\o k!}
\Big[\chi_{E+2ik}(q)-\chi_{E+2i(k+1)}(q)\Big].
}
The integral over $E$ is not well-defined in this expression.
As discussed in \KarczmarekXM, to see the behavior of 
this integral it is useful to
write $\chi_E(q)$ in terms of the non-oscillating open string character
\eqn\chiEopen{
\chi_E(q)=\rt{\pi\o2}\int_{-\infty}^\infty d\nu \,e^{\pi E\nu}
\chi_{i\nu}(q_{op}),
}
where $q$ and $q_{op}$ are the modulus in the closed string channel
and the open string channel, respectively:
\eqn\qclqop{
q=e^{-4t},\quad q_{op}=e^{-\pi^2/t}.
}
Then,  the amplitude \ZBfBinclosed\ in the open string channel is
written as  
\eqn\BfBinop{\eqalign{
&\bra B_f|q^{\hf(L_0+\til{L}_0)-{1\o24}}|B\ket\cr
=&\int{dEd\nu\o2\rt{2\pi}}{\pi^2\o\sinh^2\pi E}\chi_{i\nu}(q_{op})
\sum_{k=0}^\infty {(iE)^k\o k!}
\Big[e^{\pi\nu(E+2ik)}-e^{\pi\nu(E+2i(k+1))}\Big]\cr
=&\int{dEd\nu\o2\rt{2\pi}}{\pi^2\o\sinh^2\pi E}\chi_{i\nu}(q_{op})
(1-e^{2\pi i\nu})\exp\Big(\pi\nu E+iEe^{2\pi i\nu}\Big).
}}
The singularity at $E=0$ can be taken care of by subtracting 
the double pole $1/E^2$ \KarczmarekXM, or, by taking the pricipal value.
The factor $1-e^{2\pi i\nu}$ in \BfBinop\
vanishes when $\nu$ is integer,
which is the position of the poles in the corresponding expression of 
$\bra B_f|q^{\hf(L_0+\til{L}_0)-{1\o24}}|B_f\ket$ \KarczmarekXM.
Moreover, the large $E$ contribution is suppressed because of the oscilatory
factor $e^{iEe^{2\pi i\nu}}$.
Therefore, the incorporation of the contribution from $|B_0'\ket$
qualitatively changes the analytic structure of the amplitude.

\newsec{Discussion}
We have shown that the contribution from the
exponentially growing couplings to the timelike oscillators can
be resummed in the annulus amplitude between $|B\ket$ and $|B_f\ket$.
Although this result is suggestive, we cannot draw any conclusion about
the amplitude between two $|B\ket$'s
because we don't know the explicit form of
the off-diagonal integral $Z_{N,M}$ with $N\not=M$. 

To see whether the resummation in \ZBBasUint\ is possible,
it would be useful to find the underlying integrable structure
of grand canonical partition function as in \KazakovPM.
It would also be important to understand the
physical meaning of the appearance of the superalgebra $u(N|M)$.
In the superstring theory, the supergroup $U(N|M)$ naturally appears in the 
system of $N$ D-branes and $M$ $\b{\rm D}$-branes \VafaQF.
However, in our case, we are discussing the D-brane in the bosonic string
theory and the $N$ and $M$ are the number of insertion of tachyon operators. 
The fermionic nature of the eigenvalues in unitary matrix integral
might be related to the appearance of superalgebra.
It is also interesting to generalize our analysis to the 
boundary state of unstable D-branes in superstring theory 
along the line of \Larsen.

\vskip10mm
\centerline{\bf Acknowledgments}
I would like to thank N. Constable, M. Gutperle,
M. Jinzenji, F. Larsen, H. Liu, E. Martinec,
and M. Wakimoto for helpful discussions.
This work was supported in part by DOE grant DE-FG02-90ER40560.

\appendix{A}{Proof of \Fprodid{a,b}}
In this appendix, we give a proof of \Fprodid{a,b} 
by taking a derivative of the Jacobi's
triple product identity.
Let us first consider \Fprodid{a}. 
Notice that the summation in the r.h.s of \Fprodid{a} can be extended to the 
whole integer
\eqn\rhssum{
\sum_{k=0}^\infty(-1)^k(k+1)^2q^{k(k+2)}
=\hf\sum_{k\in\Z}(-1)^k(k+1)^2q^{k(k+2)}.
}
This sum is obtained as a limit of the following Jacoi's triple product identity
\eqn\Jacaq{\eqalign{
\sum_{k\in\Z}(-1)^ka^{(k+1)^2}q^{k(k+2)}
&=a\prod_{n=1}^\infty(1-a^{2n}q^{2n})(1-a^{2n-3}q^{2n-3})(1-a^{2n+1}q^{2n+1})\cr
&=-q^{-1}\prod_{n=1}^\infty{1-a^nq^n\o1+a^nq^n}.
}}
Here we used the well-known relation
\eqn\oddprodid{
\prod_{n=1}^\infty(1-x^{2n-1})=\prod_{n=1}^\infty{1\o1+x^n}.
}
Substituting $a=1+\vep$ into \Jacaq, and comparing the order
$\vep$ term on both sides, we have
\eqn\qtwosumid{\eqalign{
\sum_{k\in\Z}(-1)^k(k+1)^2q^{k(k+2)}&=\prod_{n=1}^\infty{1-q^n\o1+q^n}
\sum_{m=1}^\infty\lf({mq^{m-1}\o1-q^m}+{mq^{m-1}\o1+q^m}\ri)\cr
&=2\prod_{n=1}^\infty{1-q^n\o1+q^n}\sum_{k=0}^\infty{q^{2k}\o(1-q^{2k+1})^2}.
}}
Eq.\Fprodid{a} follows from \rhssum\ and \qtwosumid.

Next consider \Fprodid{b}. As in \rhssum, we first extend the range
of summation
\eqn\rhsthree{
\sum_{k=0}^\infty(-1)^k
(k+1)(k+2)(2k+3)q^{k(k+3)}=\hf\sum_{k\in\Z}(-1)^k
(k+1)(k+2)(2k+3)q^{k(k+3)}.
}
Then writing the Jacobi's identity
\eqn\Jacobabq{
\sum_{k\in\Z}(-1)^ka^{k^2}b^kq^{k(k+3)}=
\prod_{n=1}^\infty(1-a^{2n}q^{2n})(1-a^{2n-1}bq^{2n+2})
(1-a^{2n-1}b^{-1}q^{2n-4})
}
and expanding the both sides around $a=b=1$, we find
\eqn\qthreeksum{\eqalign{
\sum_{k\in\Z}(-1)^kq^{k(k+3)}&=0,\cr
\sum_{k\in\Z}(-1)^kkq^{k(k+3)}&=-q^{-2}\prod_{n=1}^\infty(1-q^{2n})^3,\cr
\sum_{k\in\Z}(-1)^kk^2q^{k(k+3)}&=3q^{-2}\prod_{n=1}^\infty(1-q^{2n})^3,\cr
\sum_{k\in\Z}(-1)^kk^3q^{k(k+3)}&=\lf(-7q^{-2}+
\sum_{k=0}^\infty{6q^{2k}\o(1-q^{2k+2})^2}\ri)\prod_{n=1}^\infty(1-q^{2n})^3.
}}
From \qthreeksum\ and \oddprodid, one can easily see that \Fprodid{b} is true.

\listrefs
\bye